\documentstyle[12pt]{article}
\begin{document}

\noindent {\Large \bf Detection of Lense-Thirring Effect Due to Earth's Spin.}
\\ \\ \\ \\

\noindent {\large \bf Ignazio Ciufolini$^{1,2}$, David Lucchesi$^{3}$,
Francesco Vespe$^{4}$, \, \, \,
\, \, \, \, and Federico Chieppa$^{5}$\normalsize \\

\noindent {1 IFSI-CNR-Frascati, Roma}

\noindent {2 Dip. Aerospaziale, Univ. Roma I "La Sapienza"}

\noindent {3 Dip. Matematica, Univ. Pisa}

\noindent {4 ASI-CGS, Matera}

\noindent {5 Scuola Ing. Aerospaziale, Univ. Roma I "La Sapienza"}

\vspace{.7in}

\noindent {\bf Rotation of a body, according to Einstein's theory of general
relativity,
generates a "force" on other matter; in Newton's gravitational theory only
the mass of a body produces a force. This phenomenon, due to currents of mass,
is known as gravitomagnetism owing to its formal analogies with magnetism due
to currents of electric charge. Therefore, according to general relativity,
Earth's rotation should influence the motion of its orbiting satellites.
Indeed, we analysed the laser ranging observations of the orbits of the
satellites LAGEOS and LAGEOS II, using a program developed at NASA/GSFC,
and obtained the first direct measurement of the gravitomagnetic orbital
perturbation
due to the Earth's rotation, known as the Lense-Thirring effect.
The accuracy of our measurement is about 25 $\%$.}

In general relativity$^{1,2}$ the concept of inertial frame has only a local
meaning
and a local inertial frame is "rotationally dragged" by mass-energy currents,
in
other words moving masses influence and change the orientation of the axes of a
local inertial frame (gyroscopes). Thus, an external current of mass, such as
the
spinning Earth, "drags" and changes the orientation of gyroscopes. This is the
"rotational dragging of inertial frames", or "frame-dragging" ("dragging", as
Einstein named it). The NASA Gravity Probe-B experiment$^{3}$ is aimed to
measure with
great accuracy this phenomenon on the orientation of the axis of spin of a
small
orbiting gyroscope. However, the whole orbital plane of a satellite is itself a
kind of enormous gyroscope dragged by the gravitomagnetic field (Figs. 1 and
2). Indeed,
in addition to the rotational dragging and precession of a test gyroscope due
to the angular momentum
${\bf J}$ of a central object, the orbit of a test particle around a central
body with angular momentum {\bf J} has a
secular rate of change of the longitude of the
line of the nodes (intersection between the orbital plane of the test particle
and the equatorial plane  of  the central object), discovered by Lense-Thirring
(1918):
$\dot {\bf \Omega\/}^{Lense-Thirring} \,  = \,{2 {\bf J}
/  [a^3 \, (1 - e^2)^{3/2}]}$; where $a$ is the semimajor axis of the test
particle, and $e$ its orbital
eccentricity. The orbit of the test particle also has a secular rate of change
of the mean
longitude of the orbit and of the longitude of the pericenter
$\dot {\bf \tilde \omega \/}$,
(defining the Runge-Lenz vector): $\dot {\bf \tilde \omega}^{Lense-Thirring} ~
= ~ {2 J  ~
( \, {\hat {\bf J}} -  3 ~ cos \, I~ \hat {\bf l} \, ) } / [ a^3  (1 - e^2)^{3
/ 2} ]$;
where $\hat {\bf l \/}$ is the orbital angular momentum, unit vector, of the
test particle,
and {\it I \/} its orbital  inclination (angle between the orbital plane
and the equatorial plane of the central  object).

Since 1896 several experiments have been discussed and proposed to measure the
rotational dragging
of inertial frames by a spinning body$^{1-4}$.
So far, the only indirect astrophysical evidence for the
rotational dragging of inertial
frames by a current of mass was given by the periastron precession rate of the
binary pulsar
PSR 1913+16$^{5}$.

Our direct measurement of the Lense-Thirring effect was obtained by laser
ranging
observations of the satellites LAGEOS and LAGEOS II$^{6}$ (Fig. 3).
The gravitomagnetic field has changed the point of closest approach to Earth --
perigee --
of the satellite LAGEOS II by about 11 meters during our period of observation
of about 3.1 years.
The semimajor axis of LAGEOS is $a \, \cong \, 12,270$ km, the
period $P \, \cong \, 3.758$ ~hr, the eccentricity $e \, \cong \, 0.004$, and
the
inclination  $I \, \cong \, 109.9^\circ$.
The semimajor axis of LAGEOS II is
$a_{II} \, \cong \, 12,163$ km, the eccentricity $e_{II} \, \cong \, 0.014$,
and the
inclination  $I_{II} \, \cong \, 52.65^\circ$.

Cornerstones of our analysis were: the NASA launch in 1976 of the LAGEOS
satellite;
the development by the NASA/Goddard Space Flight Center of the powerful program
GEODYN, and of its new version GEODYN II, for satellite orbit determination,
geodetic
parameter estimation, tracking instrument calibration, satellite orbit
prediction and other
applications in geodesy; the determination of highly accurate Earth's gravity
field solutions,
including GEML1, GEML2, GEMT1, GEMT2, GEMT3, GEMT3S. The other new basic
elements that made
our direct measurement of the Lense-Thirring effect possible were:
(1) the launch in October 1992 by NASA and ASI of the
laser-ranged satellite LAGEOS II;
(2) the new Earth's gravity field solutions$^{7}$ JGM-2 and JGM-3, jointly
developed by
NASA-Goddard and by the CSR (Center for Space Research) of the University of
Texas at Austin;
(3) the continuous laser ranging to the satellites LAGEOS and LAGEOS II from
several stations
around the world, the ranging data
from the best stations have a precision of a few millimeters; and (4) the use
of a new method$^{8}$ to
measure the gravitomagnetic field.

  We analysed the orbits of the satellites LAGEOS and LAGEOS II using existing
laser
ranging observations, a highly accurate modeling of their orbital perturbations
including the
gravity field solution JGM-3, and the 1994 version of GEODYN II. All the
general
relativistic perturbations due to the masses of Earth and Sun, including the de
Sitter or geodetic
effect (today measured with accuracy of the order of $10^{-2}$), were
incorporated in the
GEODYN equations of motion and then computer-integrated;
we did not however include in our model the orbital perturbations
due to the Earth's angular momentum, that is the Lense-Thirring,
gravitomagnetic, effect
to be determined. In order to measure the frame-dragging effect from our
residuals we introduced a
new parameter $\mu$, which, by definition, is one in general relativity,
$\mu^{GR} \equiv 1$,
and zero in Newtonian theory$^{2}$.

The residuals of the orbital elements of a satellite give a measure of any
perturbation that is not modeled
accurately enough or that is not included in the model.
The orbital elements we analysed are: the node of LAGEOS I, the node of LAGEOS
II, and the perigee
of LAGEOS II. The nodes  of LAGEOS and LAGEOS II are both dragged by the
Earth's angular
momentum; according to the Lense-Thirring formula one has:
$ \dot { \Omega}^{Lense-Thirring}_{I} \, \cong 31 \, milliarcsec/yr $
and
$ \dot { \Omega\/}^{Lense-Thirring}_{II} \,  \cong 31.5 \, milliarcsec/yr  $.
The argument of
pericenter (perigee in our analysis), $\omega$, of a test particle, that is the
angle
on its orbital plane measuring the departure of the pericenter from the
equatorial plane of the central
body, also has a Lense-Thirring drag; for LAGEOS I one has:
$  \dot { \omega \/}^{Lense-Thirring}_{I} ~ $ $ \cong ~ 32 \, milliarcsec/yr,
\, $ and for LAGEOS II:
$  \dot { \omega \/}^{Lense-Thirring}_{II} ~ $ $ \cong ~ - 57 \, milliarcsec/yr
\, $.
The nodal precessions of LAGEOS and LAGEOS II can be determined with an
accuracy of the order of
1 milliarcsec/yr, or less. In fact, we obtained a root mean square of the
node residuals of about 2 milliarcsec for LAGEOS and of about 3 milliarcsec for
LAGEOS II,
over a total period of observation of about 3.1 years. Regarding the perigee,
the observable quantity
is $e a \dot \omega$, where $e$ is the orbital eccentricity of the satellite.
Thus, for LAGEOS the perigee precession $\dot \omega$ is an extremely difficult
quantity to measure;
its orbital eccentricity is in fact about 4$\times 10^{-3}$. The orbit of
LAGEOS II is more
eccentric: its orbital eccentricity is about 0.014, and the Lense-Thirring drag
of the perigee of LAGEOS II
is almost twice as large, in magnitude, as that of LAGEOS. In fact, we obtained
a root mean square of the residuals
of the LAGEOS II perigee of about 35 milliarcsec over about 3.1 years, whereas
the
total effect of frame-dragging on the perigee, over about 3.1 years, is $\cong$
$- 176$ milliarcsec.

  The most critical source of error in our measurement arises from
uncertainties in the Earth's
  even zonal harmonics
and in their temporal variations. Using only the satellites orbiting today,
one cannot eliminate the unmodeled
orbital perturbations due to all the even zonal harmonics; in particular,
unmodeled orbital effects
due to the harmonics of lower order are of a size comparable to or larger than
the Lense-Thirring effect.
However, by analysing the JGM-3 solution with its uncertainties in the even
zonal harmonic
coefficients, and by calculating the secular effects of these uncertainties on
the orbital elements
of LAGEOS and LAGEOS II, we found that the main sources of error in the
determination
the frame-dragging effect
are concentrated in the first 2 or 3 even zonal harmonics, that is $J_{2}$,
$J_{4}$ and $J_{6}$.
To further test$^{10}$ the order of magnitude of
the real errors in the estimated value of the $J_{2n}$
coefficients, we took the
difference$^{8}$ between two different gravity field solutions: JGM-3 and
GEMT-3S$^{9}$.
We then found that by far the largest uncertainties, on the nodes of LAGEOS and
LAGEOS II and on the perigee of
LAGEOS II, arise from $\Delta J_{2}$ and $\Delta J_{4}$, a smaller error is due
to  $\Delta J_{6}$,
and much smaller errors arise from the differences in the other $J_{2n}$
coefficients. However, we have
the three observable quantities: the node of LAGEOS, the node of LAGEOS  II,
and the perigee
of LAGEOS II, and we want to determine the parameter $\mu$, measuring the
frame-dragging
effect. Then, we can use these three observable quantities
$\dot \Omega_{I}$, $\dot \Omega_{II}$ and $\dot \omega_{II}$ to determine $\mu$
thereby eliminating the
two largest sources of error arising from the uncertainties in $J_{2}$ and
$J_{4}$.
This new method leads to a value of $\mu$ unaffected by the errors due to
$\delta J_{2}$
and $\delta J_{4}$, by far the largest, but sensitive only to the smaller
errors due to
$\delta J_{2n}$ with $2n \geq 6$.
As regards tidal, secular and seasonal changes in the geopotential
coefficients,
we stress that the main effects on the
nodes and perigee of LAGEOS and LAGEOS II due to tidal and other temporal
variations
in the Earth's gravity field are due to changes in the first two even zonal
harmonic coefficients, $ J_{2}$ and $ J_{4}$. Any tidal error in $ J_{2}$ and $
J_{4}$,
and any error due to other unmodeled temporal variations in $ J_{2}$ and $
J_{4}$,
including their secular and seasonal variations, is eliminated using our
combination of
residuals of nodes and perigee. In particular, most of the errors due to the
18.6 year and
9.3 year tides, associated
with the Moon node, are eliminated in our measurement.
Thus, using three observable quantities, the two nodes and the perigee, one can
solve
for $\mu$ and eliminate
$\delta J_{2}$ and $\delta J_{4}$:
$ { \delta \, \dot \Omega^{Exp}_{LageosI} + k_{1} \,
\delta \, \dot \Omega^{Exp}_{LageosII} + k_{2} \, \delta \, \dot
\omega^{Exp}_{LageosII} \,} $ = $ {\,
\mu  \, ( \, 31 \, + \, 31.5 \, k_{1} \, - \, 57 \, k_{2}) \, milliarcsec/yr} $
$ {+ ~ [ contributions ~ from ~ \delta J_{6}, \delta J_{8}, ... \, ], } $
where $k_{1} = 0.295$ and $k_{2} = - 0.35$ are obtained (in order to eliminate
the $\delta J_{2}$ and $\delta J_{4}$ errors) from the system of the three
equations
for the nodal rates of LAGEOS and LAGEOS II and for the perigee rate of LAGEOS
II.
The best fit lines of the residuals of the nodes  of LAGEOS and LAGEOS II had a
slope of
respectively $\cong - 11$ milliarcsec and $\cong$ 40 milliarcsec, and
the best fit line of the residuals of the perigee of LAGEOS II had a slope of
$\cong -$ 188 milliarcsec.
In Fig. 4 we plotted the sum of the
residuals of the nodes of LAGEOS and LAGEOS II and
perigee of LAGEOS II
according to our formula to eliminate the $\delta J_{2}$ and $\delta J_{4}$
errors
and after having removed 10 small periodical residual signals (corresponding to
9 main
tidal effects and to the largest solar radiation pressure perturbation)
and the small observed inclination residuals.
In other words each point of Fig. 4  was obtained by one residual of the node
of LAGEOS, plus
the corresponding residual of the node of LAGEOS II times the factor $0.295$,
plus
the corresponding residual of the perigee of LAGEOS II times the factor $-
0.35$.
By fitting a straight line through these combined residuals of nodes and
perigee
(obtained using the JGM-3 gravity field model) we finally found:

 {\begin{equation} \mu \cong 1.1 \; ,  \end{equation}}

This combined, measured, gravitomagnetic perturbation
of the satellites' orbits corresponds to about 12 meters at the LAGEOS
altitude,
that is about 205 milliarcsec. The root mean square of the post-fit combined
residuals is
about 13 milliarcsec.
The main error sources affecting the nodes of LAGEOS and LAGEOS II and the
perigee of LAGEOS II
are: errors due to uncertainties in the even zonal harmonics, $J_{2n}$ (with $
2n \geq 6$ in our
measurement);
errors due to unmodeled tidal perturbations and other temporal variations in
the Earth's gravity field
(due to ${\dot J}_{2n}$ with $ 2n \geq 6$ in our measurement),
random and stochastic observational errors;
errors due to uncertainties in the orbital inclinations (though we corrected
nodes and perigee
with the residuals of the orbital inclinations); errors due
to nongravitational perturbations, including direct solar radiation pressure,
Earth's albedo,
Yarkovsky anisotropic thermal radiation, Rubincam effect (anisotropic
re-radiation of Earth infrared
radiation absorbed by the LAGEOS retro-reflectors), particle drag, and errors
due to the estimated
values of the satellite's reflectivities and estimated 15-day along track
accelerations.

By calculating the effects of
all these systematic and random error sources (paper in preparation), we found:
\begin{equation}   \delta \mu  {\buildrel < \over \sim} ~ 25 \, \% \, \mu
\end{equation}
\noindent
In conclusion, we obtained the result:

               \begin{equation}    \mu = 1.1 \pm 0.25  \end{equation}

       \noindent   (whereas $\mu$ = 1 in general relativity).

\vspace{.25in}

\begin{enumerate}

\item  Misner, C.W., Thorne, K.S., and Wheeler, J.A. {\it Gravitation}
(Freeman, San
Francisco, 1973).

\item  Ciufolini, I., and Wheeler, J.A. {\it Gravitation and Inertia}
(Princeton
University Press, Princeton, New Jersey, 1995).

\item Everitt, C. W. F., in {\it Experimental Gravitation} (ed Bertotti B.)
331-360
(Academic Press, New York, 1974).

\item Ciufolini, I. {\it Phys. Rev. Lett.} {\bf 56\/}, 278-281 (1986).

\item Nordtvedt K. {\it Int. J. Theoret. Phys.} {\bf 27\/}, 1395-1403 (1988).

\item Cohen, S. C. et al., ed., {\it LAGEOS Scientific Results, J. Geophys.
Res.,\/}  {\bf 90\/} (B11),
9215-9438 (1985).

\item Nerem, R. S. et al. {\it J. Geophys. Res.,\/} {\bf 99\/} (C12),
24,421-24,447 (1994).

\item Ciufolini, I. et al. {\it Nuovo Cimento A}, (1986).

\item Lerch, F. J. et al. {\it J. Geophys. Res.,\/} {\bf 99\/} (B2), 2815-2839
(1994).

\item Lerch, F. J., Klosko, S. M., Wagner, C. A., and Patel, G. B.
{\it J. Geophys. Res.,\/} {\bf 90\/} (B11), 9312-9334 (1985).

\item Thorne K. S. et al., ed., {\it Black Holes, the Membrane Paradigm} (Yale
University Press,
New Haven and London, 1986).

\end{enumerate}

\vspace{.25in}

\noindent ACKNOWLEDGEMENTS. This work was significantly aided by several
programs and facilities
of NASA in particular, through data provided to us by the CDDIS of the
NASA Goddard Space Flight Center and the supply to us of the program GEODYN II.
We also appreciate fundamental help
from Steve Klosko, Kenneth Nordtvedt, Erricos Pavlis and Mark Torrence.

\vspace{8in}

\noindent FIG. 1  The gravitomagnetic field, $\bf{H}$, generated by the angular
momentum, $\bf{J}$, of a central rotating body. In general relativity,
for a localized, stationary,
mass-energy distribution, in the weak-field and slow-motion limit,
the three "vector" components of the metric tensor are given by:
${\bf{h}}\cong -2 (\bf{J}\times
\bf{x})/{r^3}$, where $\bf{J}$ is the angular momentum of
the central body and ${\bf{h}}$ is known as the gravitomagnetic potential.
The gravitomagnetic field $\bf{H}$ is given by $\bf{H}%
=\bf{\nabla }\times \bf{h.}$ To characterize the gravitomagnetic field
generated by the angular
momentum of a body
and the Lense-Thirring effect, and distinguish it from other relativistic
phenomena such
as the de Sitter effect -- due to the motion of a gyroscope in a static
gravitational field --
one may give a description of the gravitomagnetic field in terms of
spacetime-curvature invariants.
The pseudoinvariant $ ^\ast {\bf R\/}\cdot {\bf R\/ }$, built from the Riemann
tensor ${\bf R}$ and
its dual $ ^\ast {\bf R}$, gives an invariant characterization of
gravitomagnetism since it is nonzero
in the field of a central body if and only if the body is rotating. Indeed the
pseudoinvariant
$ ^\ast {\bf R\/}\cdot {\bf R\/ }$ is proportional to
the angular momentum of the central body.
Thus, one may describe gravitomagnetism as that phenomenon of nature such that
{\it spacetime
curvature is generated by the spin of a body}$^{2}$. \\

\noindent FIG. 2  A twisted jet from the nucleus of the galaxy 3C 66B. This
ultraviolet picture
has been taken by the ESA Faint Object Camera of the NASA Hubble Space
Telescope. The twisted jet
of plasma extends 10,000 light-years from the nucleus of the galaxy 3C 66B
located at about 270
million light-years from Earth. The ultraviolet radiation is emitted by
electrons in the jet
spiraling through magnetic fields. Long radio jets from quasars and active
galactic nuclei are
observed to have constant directions in space, which correspond to emission
time scales that may
reach millions of years. The constant direction of the jets suggests the
existence of a central
astrophysical gyroscope; this engine and gyroscope might be a super-massive
spinning black
hole with its gravitomagnetic field. The constant orientation of the emitted
jets may then
be explained$^{11}$ using the gravitomagnetic field of the central spinning
body.\\

\noindent FIG. 3 The laser-ranged satellite LAGEOS II.
Laser ranging to the Moon and to artificial satellites is an impressive
technique to
measure distances from a laser-tracking station on Earth to retro-reflectors
placed on the Moon, or on
satellites orbiting Earth. By the use of short laser pulses ranges can be
measured with accuracies of
less than 1 cm from emitting lasers on Earth to retro-reflectors on a
satellite, and with accuracies of
less than 10 cm to retro-reflectors on the Moon.  The NASA-ASI (Italian Space
Agency) satellite LAGEOS II
is a high-altitude, small cross-sectional
area-to-mass ratio, spherical, laser-ranged satellite. It is made
of heavy brass and aluminum, is completely passive and
covered with laser retro-reflectors. It acts as a reference target for
ground-based laser-tracking systems to measure -- via laser ranging --  crustal
movements, plate
motion, polar motion and Earth rotation.
LAGEOS II is essentially identical to the NASA satellite LAGEOS (LAser
GEOdynamics Satellite)
but they have different orbital parameters. \\

\noindent FIG. 4  Sum of the residuals of the nodes of LAGEOS and LAGEOS II and
perigee
of LAGEOS II from November 1992 to December 1995, using the method described in
the text.
 On the vertical axis we plotted $(node \, residuals \, of \, LAGEOS) \,
+ \, 0.295 \, (node \, residuals \, of \, LAGEOS II) \, -$

\noindent $- \, 0.35 \, (perigee \,
residuals \, of \, LAGEOS II)$.
In our analysis we included polar motion from VLBI (IERS), Earth's solid and
ocean tides
and Earth's gravity field, GM and spherical harmonics up to order 50, from the
JGM-3
gravity field model, solar, lunar
and planetary perturbations and nongravitational perturbations including solar
radiation pressure,
Earth's radiation pressure, anisotropic thermal radiation effects, and
atmospheric drag.
For each 15-day arc we estimated all station coordinates except the latitude
of Goddard Space Flight Center and the latitude/longitude of Hawaii
(maintained fixed), the spacecrafts' initial conditions
(initial positions and velocities), the
satellites' reflectivities and 15-day along-track accelerations.
The best fit line shown through these combined residuals has a slope of about
66 milliarcsec/yr
(the total integrated effect corresponds to about 12 meters at the LAGEOS
altitude),
that is $\mu \cong 1.1$
(whereas $\mu \equiv 1$ in general relativity), and the corresponding root mean
square
of the residuals is about 13 milliarcsec.
Due to systematic (secular and periodical) errors and random errors,
we estimated the total error in our measurement of $\mu$
to be less than 25$\%$ of $\mu$. \\

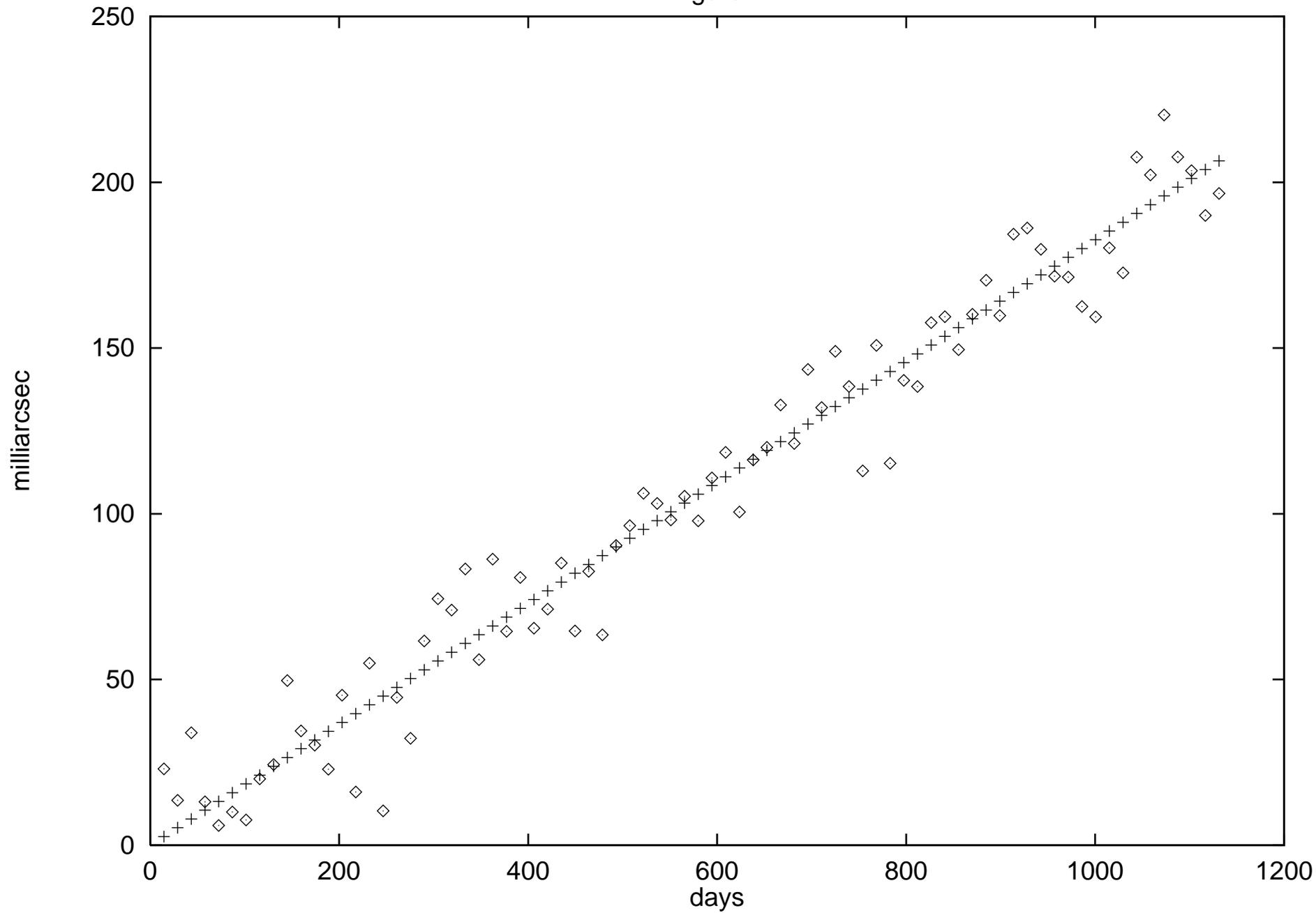
\end{document}